# Well-resolved velocity fields using discontinuous Galerkin shallow water solutions

Janice Lynn Ayog[1]; Georges Kesserwani, Ph.D[2]; and Domenico Baú, Ph.D[3]

[1]Ph.D Candidate, Department of Civil and Structural Engineering, University of Sheffield, Mappin St, Sheffield City Centre, Sheffield S1 3JD, UK. Email: jlayog1@sheffield.ac.uk.

[2]Research Fellow and Senior Lecturer, Department of Civil and Structural Engineering, University of Sheffield, Mappin St, Sheffield City Centre, Sheffield S1 3JD, UK. Email: g.kesserwani@sheffield.ac.uk (corresponding author).

[3]Senior Lecturer, Department of Civil and Structural Engineering, University of Sheffield, Mappin St, Sheffield City Centre, Sheffield S1 3JD, UK. Email: d.bau@sheffield.ac.uk.

**Abstract**

Computational models based on the depth-averaged shallow water equations (SWE) offer an efficient choice to analyse velocity fields around hydraulic structures. Second-order finite volume (FV2) solvers have often been used for this purpose subject to adding an eddy viscosity term at sub-meter resolution, but have been shown to fall short of capturing small-scale field transients emerging from wave-structure interactions. The second-order discontinuous Galerkin (DG2) alternative is significantly more resistant to the growth of numerical diffusion and leads to faster convergence rates. These properties make the DG2 solver a promising modelling tool for detailed velocity field predictions. This paper focuses on exploring this DG2 capability with reference to an FV2 counterpart for a selection of test cases that require well-resolved velocity field predictions. The findings of this work lead to identifying a particular setting for the DG2 solver that allows for obtaining more accurate and efficient depth-averaged velocity fields incorporating small-scale transients.

**Introduction**

Detailed spatial distribution of velocity is useful to support hydraulic engineering designs and applications, such as transport analyses for predicting the accumulation of woody debris around bridge piers (Mazur et al. 2016; Pagliara and Carnacina 2013; Ruiz-Villanueva et al. 2017) and the distribution of suspended sediments and solute particles in large river system (Dinehart and Burau 2005; Jodeau et al. 2008; Legleiter and Kinzel 2020). To obtain detailed velocity fields, direct in-situ measurement is not often viable due to safety concerns and lack of access to deploy surveying equipment. Instead, the velocity fields are analysed in physical models that are designed to replicate real-world hydrodynamics, which is useful when analysing complex wave features around specific engineering structures. For instance, physical models have been used to study the recirculation patterns of velocity fields in channel lateral cavities (Jackson et al. 2015; Juez et al. 2018; Mignot and Brevis 2020), diverting junctions (Mignot et al. 2013; Momplot et al. 2017; Shettar and Murthy 1996) and around dense building blocks in urban areas (Chen et al. 2018; LaRocque et al. 2013; Smith et al. 2016). As complex physical models tend to be costly, computational models can be used as a direct alternative to them, or support the analysis of data from physical experiments.

In order to avoid the prohibitive cost of three-dimensional (3D) Navier-Stokes computational models, numerical solvers of the 2D depth-averaged shallow water equations (SWE) have been used instead to produce sufficiently detailed velocity fields at a viable runtime for field-scale hydraulic engineering applications (Aureli et al. 2015; Duan 2005). Such 2D-SWE solvers are often build around second-order finite volume (FV2) numerical methods (Wang et al. 2013), which are also featured in industry-scale hydraulic modelling packages, such as Rubar20 (INRAE 2021), Iber (Bladé et al. 2014) and TUFLOW-HPC (BMT-WBM 2018). Despite their wide application to model the velocity fields around areas of wave-structure interactions, FV2 solvers have been reported to require the addition of an eddy viscosity term as the resolution becomes finer and when there is drastic change in the velocity magnitude and direction (Collecutt and Syme 2017; Syme 2008). Adding the eddy viscosity term can reduce the FV2 solver tendency to over-expand the extent of the recirculation flows and, thereby, improves the modelling of spatial velocity profiles (Bazin 2013). However, the eddy viscosity approach requires case-dependent and often quite onerous calibration efforts. In addition, the fast growth rate of numerical diffusion in FV2 solvers has been observed to smear the presence of small-scale eddies in the recirculation zones (Cea et al. 2007; Özgen-Xian et al. 2021), leading to significant discrepancies in the velocity field predictions around hydraulic structures such as bridge piers (Horritt et al. 2006).

The second-order discontinuous Galerkin (DG2) solver, compared to the FV2 solver, has a significantly lower rate of numerical diffusion, as demonstrated in the comparative study by Ayog et al. (2021). This property enables the DG2 solver of the SWE to capture small-scale eddies in velocity fields, without the need of an eddy viscosity term, as also hinted in a few published papers addressing coastal and estuary environments. Kubatko et al. (2006) studied the performance of DG2 and higher-order DG solvers in reproducing 2D velocity fields for a tidal flow over an idealised channel. The investigators concluded that the DG2 solver can deliver well-captured eddies outside of the channel inlet, leading to velocity fields that are as accurate as those produced by higher-order DG solvers. Alvarez-Vázquez et al. (2008) applied a DG2 solver for the simulation of fish migration to support the design of vertical slots along a river fishway structure, with results showing a great potential for the DG2 solver to reproduce small-scale eddies within its 2D velocity field predictions. Beisiegel et al. (2020) explored a DG2 solver to simulate flow circulation, reporting that their DG2 solver was able to replicate the asymmetrical patterns of the recirculation eddies extracted from a 3D model. Since these papers only hint at the potential benefits of DG2 solvers to produce detailed velocity fields for coastal and ecological applications, a dedicated investigation is needed to more closely explore this aspect for smaller-scale hydraulic engineering applications that primarily require detailed velocity field prediction.

This paper investigates a grid-based DG2 solver that has been initially designed for practical flood modelling applications by Ayog et al. (2021). The capabilities of this DG2 solver are studied in relation to spatial velocity fields with recirculation zones through a series of selected test cases in which spatial velocity field experiments are available. In the next section, the DG2 solver is overviewed alongside the standard FV2 solver, and the selected test cases are presented. These test cases involve small-scale flow interaction with steep topographic structures, for a wide range of flow regimes and transitions and considering a real-world flooding scenario in a residential area. The performance of the DG2 solver is then analysed for two configurations, with and without activating local slope limiting, jointly with that of the FV2 solver. Analysis of performances include a qualitative comparison of predicted velocity streamlines in the areas with recirculation flows, and an evaluation based on a range of quantitative indices. Conclusions from the analyses are finally drawn to come up with a most suited configuration for the DG2 solver for more efficient and accurate 2D velocity field predictions, which are detailed enough to capture the small-scale features in and around the zones of wave-structure interactions.

**Materials and Methods**

The DG2 solver adopts the "slope-decoupled" simplification approach of Kesserwani et al. (2018), which can be formulated on a calculation stencil similar to that of grid-based FV-based solvers (Ayog et al. 2021). Compared to a standard DG2 solver formulated upon a full tensor product stencil, the slope-decoupled DG2 solver is 2.6 times more efficient to run and preserves second-order mesh convergence and the robustness properties of FV-based solvers for handling irregular topography with wetting and drying (Kesserwani et al. 2018). When applied to simulating gradually propagating flood flows at a modelling resolution larger than 5 m, the slope-decoupled DG2 solver is robustly applicable without local slope limiting in favour of doubling runtime efficiency. Activating local slope limiting is required in those cases where there are rampant noises appearing in the DG2 solutions that need to be removed, particularly in areas with strong waves reflected against steep-sloped structures and where highly resolved modelling outputs at a sub-meter resolution are sought (Ayog et al. 2021).

In the following, the DG2 solver discretisation is briefly overviewed for its two variants with local slope limiting and with no limiting, named hereafter DG2-LL and DG2-NL, respectively. These DG2 variants have been compared for industry-scale flood modelling benchmark test cases and are reconsidered to further assess their capabilities for smaller-scale simulations involving high-resolution spatial velocity fields. An FV2 solver using the Monotonic Upstream-centred Scheme for Conservation Laws (MUSCL-FV2) is also considered in order to compare the differences among the MUSCL-FV2 and DG2 solvers in terms of velocity prediction capabilities. A detailed technical description of the DG2 and MUSCL-FV2 solvers can be found in Ayog et al. (2021), and the corresponding parallelised codes can be accessed from the University of Sheffield local repository of the LISFLOOD-FP8.0 modelling packages on GitHub (Shaw et al. 2021; University of Sheffield 2021). Therefore, the DG2 and MUSCL-FV2 solvers used to support this investigation are only briefly overviewed next.

*Overview of the grid-based DG2 and MUSCL-FV2 solvers*

The discrete formulations of the DG2 and MUSCL-FV2 solvers for the numerical solution to the SWE follow the Godunov-type framework (Toro 2001; Toro and Garcia-Navarro 2007), without any added eddy viscosity and/or turbulence term. The SWE equations can be presented in a conservative matrix form:

$$\partial_t \mathbf{U} + \partial_x \mathbf{F}(\mathbf{U}) + \partial_y \mathbf{G}(\mathbf{U}) = \mathbf{S}(\mathbf{U}) \tag{1}$$

where $\partial$ represent a partial derivative operator, $\mathbf{U}(x, y, t) = [h, h \cdot u, h \cdot v]^T$ is the flow vector at time *t* and location (*x*, *y*) and involves the water depth *h* (m), the depth-averaged Cartesian components of the velocity field *u* and *v* (ms$^{-1}$), and the discharges per unit width *h·u* (m$^2$s$^{-1}$) and *h·v* (m$^2$s$^{-1}$); $\mathbf{F} = [hu, (hu)^2 h^{-1} + 0.5gh^2, huv]^T$ and $\mathbf{G} = [hv, huv, (hv)^2 h^{-1} + 0.5gh^2]^T$ are vectors representing the components of physical flux field, and *g* is the gravity acceleration (ms$^{-2}$). $\mathbf{S}$ represents the source term vector including topography gradients and friction effects based on the Manning's roughness coefficient, $n_M$ (m$^{1/3}$s$^{-1}$). With both DG2 and MUSCL-FV2 solvers, Eq. (1) is solved on a 2D domain that is discretised into a number *M* of square grid $Q_c$ (*c* = 1, ..., *M*), with each grid centered at ($x_c$, $y_c$) and having the size *Δx = Δy*.

The DG2 solver uses local piecewise-planar solutions on each grid element $Q_c$ to approximate $\mathbf{U}$ as $\mathbf{U_h}$ that is generated using an average coefficient, $\mathbf{U}_c^0(t)$, and two directionally-independent slope coefficients, $\mathbf{U}_c^{1x}(t)$ and $\mathbf{U}_c^{1y}(t)$:

$$\mathbf{U_h}(x, y, t)|_{Q_c} = \mathbf{U}_c^0(t) + \frac{(x - x_c)}{\Delta x/2} \mathbf{U}_c^{1x} + \frac{(y - y_c)}{\Delta y/2} \mathbf{U}_c^{1y} \tag{2}$$

The coefficients $\mathbf{U}_c^0(t)$, $\mathbf{U}_c^{1x}(t)$ and $\mathbf{U}_c^{1y}(t)$ are evolved in time within a two-stage Runge-Kutta (RK2) integration approach whilst using three DG2 spatial operators $\mathbf{L}_c^0$, $\mathbf{L}_c^{1x}$ and $\mathbf{L}_c^{1y}$:

$$\mathbf{L}_c^0 = -\frac{1}{\Delta x}\left(\widetilde{\mathbf{F}}_E - \widetilde{\mathbf{F}}_W\right) - \frac{1}{\Delta y}\left(\widetilde{\mathbf{G}}_N - \widetilde{\mathbf{G}}_S\right) + \mathbf{S}(\mathbf{U}_c^0(t)) \tag{3a}$$

$$\mathbf{L}_c^{1x} = -\frac{3}{\Delta x}\left\{\left(\widetilde{\mathbf{F}}_E + \widetilde{\mathbf{F}}_W\right) - \left(\mathbf{F}\left(\mathbf{U}_c^0(t) + \frac{1}{\sqrt{3}}\mathbf{U}_c^{1x}(t)\right) + \mathbf{F}\left(\mathbf{U}_c^{0x}(t) - \frac{1}{\sqrt{3}}\mathbf{U}_c^{1x}(t)\right)\right) - \frac{\Delta x \sqrt{3}}{6}\left[\mathbf{S}\left(\mathbf{U}_c^0(t) + \frac{1}{\sqrt{3}}\mathbf{U}_c^{1x}(t)\right) - \mathbf{S}\left(\mathbf{U}_c^0(t) - \frac{1}{\sqrt{3}}\mathbf{U}_c^{1x}(t)\right)\right]\right\} \tag{3b}$$

$$\mathbf{L}_c^{1y} = -\frac{3}{\Delta y}\left\{\left(\widetilde{\mathbf{G}}_N + \widetilde{\mathbf{G}}_S\right) - \left(\mathbf{G}\left(\mathbf{U}_c^0(t) + \frac{1}{\sqrt{3}}\mathbf{U}_c^{1y}(t)\right) + \mathbf{G}\left(\mathbf{U}_c^0(t) - \frac{1}{\sqrt{3}}\mathbf{U}_c^{1y}(t)\right)\right) - \frac{\Delta y \sqrt{3}}{6}\left[\mathbf{S}\left(\mathbf{U}_c^0(t) + \frac{1}{\sqrt{3}}\mathbf{U}_c^{1y}(t)\right) - \mathbf{S}\left(\mathbf{U}_c^0(t) - \frac{1}{\sqrt{3}}\mathbf{U}_c^{1y}(t)\right)\right]\right\} \tag{3c}$$

After implementing wetting and drying treatments, the inter-elemental fluxes $\widetilde{\mathbf{F}}_E$, $\widetilde{\mathbf{F}}_W$, $\widetilde{\mathbf{G}}_N$ and $\widetilde{\mathbf{G}}_S$ are evaluated using the HLL approximate Riemann solver (Kesserwani et al. 2008; Toro 2001) applied to the two limits of the piecewise-planar solutions (Eq. 3a-c) from the eastern, western, northern and southern faces of the element grid $Q_c$. The DG2 solver computes the limits using the locally-available slope coefficients. For example, at the eastern face, its solution limits are computed using the *x*-directional slope coefficient, $\mathbf{U}^{1x}$, such that: $\mathbf{U}_E^- = \mathbf{U}_c^0(t) + \mathbf{U}_c^{1x}(t)$ and $\mathbf{U}_E^+ = \mathbf{U}_{nei_E}^0(t) - \mathbf{U}_{nei_E}^{1x}(t)$, where $nei_E$ is the index of the average and slope coefficients located at the neighbouring element from the east. Local slope limiting is only required to control very steep slope coefficients that would induce spurious oscillations in the DG2 solution. Local slope limiting is applied using Krivodonova et al. (2004)'s method, which allows to detect where the generalised minmod limiter (Cockburn and Shu 2001) needs to be applied.

The FV2 solver only uses a local piecewise-constant solution in each element $Q_c$, $\mathbf{U}_c^0(t)$, which is also evolved by the RK2 scheme based only on the spatial operator $\mathbf{L}_c^0$ (Eq. 3a). The inter-elemental flux evaluations are also achieved by the HLL Riemann solver but are applied to the limits reconstructed from MUSCL piecewise-linear interpolations. For example, the solution limits at the eastern face are computed as $\mathbf{U}_E^- = \mathbf{U}_c^0(t) + 0.5\nabla_c^{1x}$ and $\mathbf{U}_E^+ = \mathbf{U}_{nei_E}^0(t) - 0.5\nabla_{nei_E}^{1x}$, where the $\nabla_c^{1x}$ is the *x*-directional slope term reconstructed from the piecewise-constant solutions:

$$\nabla_c^{1x} = \phi\left(\frac{\mathbf{U}_c^0(t) - \mathbf{U}_{nei_W}^0(t)}{\mathbf{U}_{nei_E}^0(t) - \mathbf{U}_c^0(t)}\right)\left(\mathbf{U}_c^0(t) - \mathbf{U}_{nei_W}^0(t)\right) \quad (4)$$

The function $\phi$ is a global slope limiter function intrinsic to the MUSCL approach, which allows to achieve second-order and oscillatory-free numerical solutions. The symmetric limiter function of Sweby (1984) with the β parameter value of 1.25 is used, based on the sensitivity analysis in Ayog et al. (2021).

*Selection of test cases and quantitative indices*

Test cases with experimental velocity fields are selected to study the predictive capability of the DG2 solver in the reproduction of spatial 2D velocity fields, at the sub-meter resolution, for hydraulic simulation problems characterized by wave-structure interactions. The features of the selected test cases are summarised in Table 1. In the first two tests, the DG2-LL and DG2-NL variants are compared to identify and adopt a DG2 configuration that is most suited for the intended hydraulic simulation problems. The adopted DG2 variant is analysed further in the other test cases, all of which involve alternative simulation outcomes from the MUSCL-FV2 predictions to investigate the transferability of the DG2 solver to practical-scale velocity fields.

**Table 1.** The description of the selected test cases and their respective features.

| Test case | Features |
| --- | --- |
| Dam-break flow through multiple blocks | Highly transient flow narrowly constricted between sharp-edged obstacles followed by a downstream hydraulic jump. |
| Flow separation at a T-junction | Steady subcritical flow separation considering two different T-junction configurations - with and without a small obstacle – with a recirculation zone along the lateral branch. |
| Quasi-steady flow in sharp building cavities | Inflow with skewed velocities, and high-resolution measurements of 2D velocity fields containing small-scale eddies in the side cavities of two adjacent buildings. |
| Flooding in an urban residential area | Real-world flash flooding scenario around buildings on small piers simulated using a high-resolution digital elevation model (DEM). |

The 2D velocity vectors and spatial profiles produced by DG2 and MUSCL-FV2 solvers are quantitatively analysed using two indices, the $R^2$ coefficient of determination and the $L^1$-norm error. These indices are expressed as:

$$R^2 = \left[\frac{\sum_{k=1}^{N_s}(V_k^{EXP} - \bar{V}^{EXP})(V_k^{NUM} - \bar{V}^{NUM})}{\sqrt{\sum_{k=1}^{N_s}(V_k^{EXP} - \bar{V}^{EXP})^2 \sum_{k=1}^{N_s}(V_k^{NUM} - \bar{V}^{NUM})^2}}\right] \quad (5)$$

$$L^1 - \text{norm error} = \frac{1}{N_s}\left(\sum_{k=1}^{N_s}|V_k^{EXP} - V_k^{NUM}|\right) \quad (6)$$

In Eqs. (5-6), $V^{EXP}$ and $V^{NUM}$ are the velocity magnitudes from the experiments, whereas $\bar{V}^{EXP}$ and $\bar{V}^{NUM}$ represent their numerical counterparts calculated as space-averaged velocity magnitudes. $N_s$ denotes the total number of sampled data ($k = 1, … N_s$). The indices for the spatial profiles are calculated in a similar manner to the 2D velocity vectors, but by using the spatial point values instead of the velocity magnitudes. The R² coefficient can vary between 0 and 1, and quantifies the statistical correlation, or similarity, between the numerically-predicted profiles or velocity magnitudes and those from the experiments. On the other hand, the L¹-norm error provides an estimate of the discrepancy between the simulated values and the experimental data. Besides these indices, another quantitative index, named the relevance index (RI), is used to analyse velocity field directions, and is expressed as:

$$\text{RI} = \frac{1}{N_s}\left(\sum_{k=1}^{N_s}\frac{(u_k^{EXP} \times u_k^{NUM}) + (v_k^{EXP} \times v_k^{NUM})}{V_k^{EXP}V_k^{NUM}}\right) \quad (7)$$

where $u^{EXP}$ and $v^{EXP}$ are the 2D velocity components from experimental observations and $u^{NUM}$ and $v^{NUM}$ are those from the numerical solvers. The coefficient RI is used to quantify the directional alignment of the predicted velocity fields to experimental observation, with the velocity alignment becoming higher as the RI approaches unity.

**Results and Discussions**

***Dam-break flow through multiple blocks***

This test is aimed to identify a range of practical applications at which local limiting within DG2 (DG2-LL) is needed. The DG2-LL has been shown to be as reliable as the DG2 variant with no limiting (DG2-NL) for industrial-scale flood modelling applications, although twice as expensive to run (Ayog et al. 2021). Only when analysing the hydrodynamics of a dam-break wave interaction around one building block, the DG2-LL proved necessary to capture physically-meaningful depth and velocity time histories at fixed spatial points behind the block (Ayog et al. 2021). Therefore, a further investigation is required on the validity of the DG2-LL variant in a more challenging setting, such as when extracting depth and velocity spatial profiles at a given time and when the dam-break flow hydrodynamics is affected by the presence of multiple building blocks.

The test is based on the experiment of Soares-Frazao and Zech (2008). It consists of a long flume (Fig. 1) with a smooth bed surface ($n_M$ = 0.01 m$^{1/3}$s$^{-1}$). The flume contains twenty-five square blocks, each of 0.3 m × 0.3 m dimension and separated by a 0.1 m gap. The blocks are meant to represent an idealised city located on a flat floodplain downstream of the flume, which also contains a gate separating a 0.4 m and a 0.011 m deep water body. When the gate is swiftly removed, a dam-break wave occurs to propagate toward the blocks and interact with them. The presence of the blocks on the floodplain constricts the flow from moving further into the narrow gaps, except for a small amount of water that accelerates across the blocks and forms a hydraulic jump downstream. The constricted flow forms a backwater zone in which the water depth upstream of the blocks is raised. The backwater zone spreads to the flume side walls over time and pushes lateral flows between the building blocks until the flow stabilises. This simulation is run up to $t = 10$ s on a mesh with a grid size of 0.02 m. The longitudinal profiles of water depth and velocity obtained from DG2-NL, DG2-LL and MUSCL-FV2 at $t =$ 4, 5, 6 and 10 s are extracted along $y = 0.2$ m (red line, Fig. 1). They are shown in Fig. 2, where they are also compared with observed longitudinal profiles.

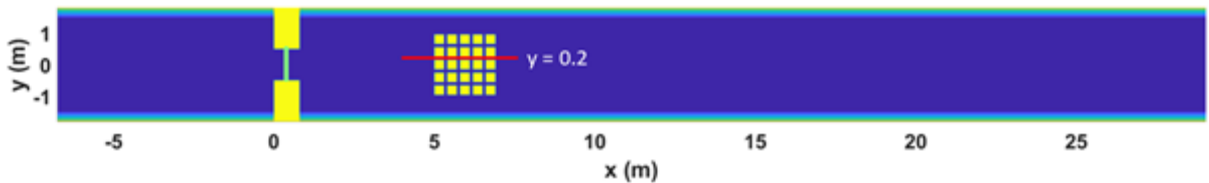

**Fig. 1.** Plan view of the spatial domain, indicating the location of the gate (green line) and the location of the extracted water level and velocity profiles (red line).

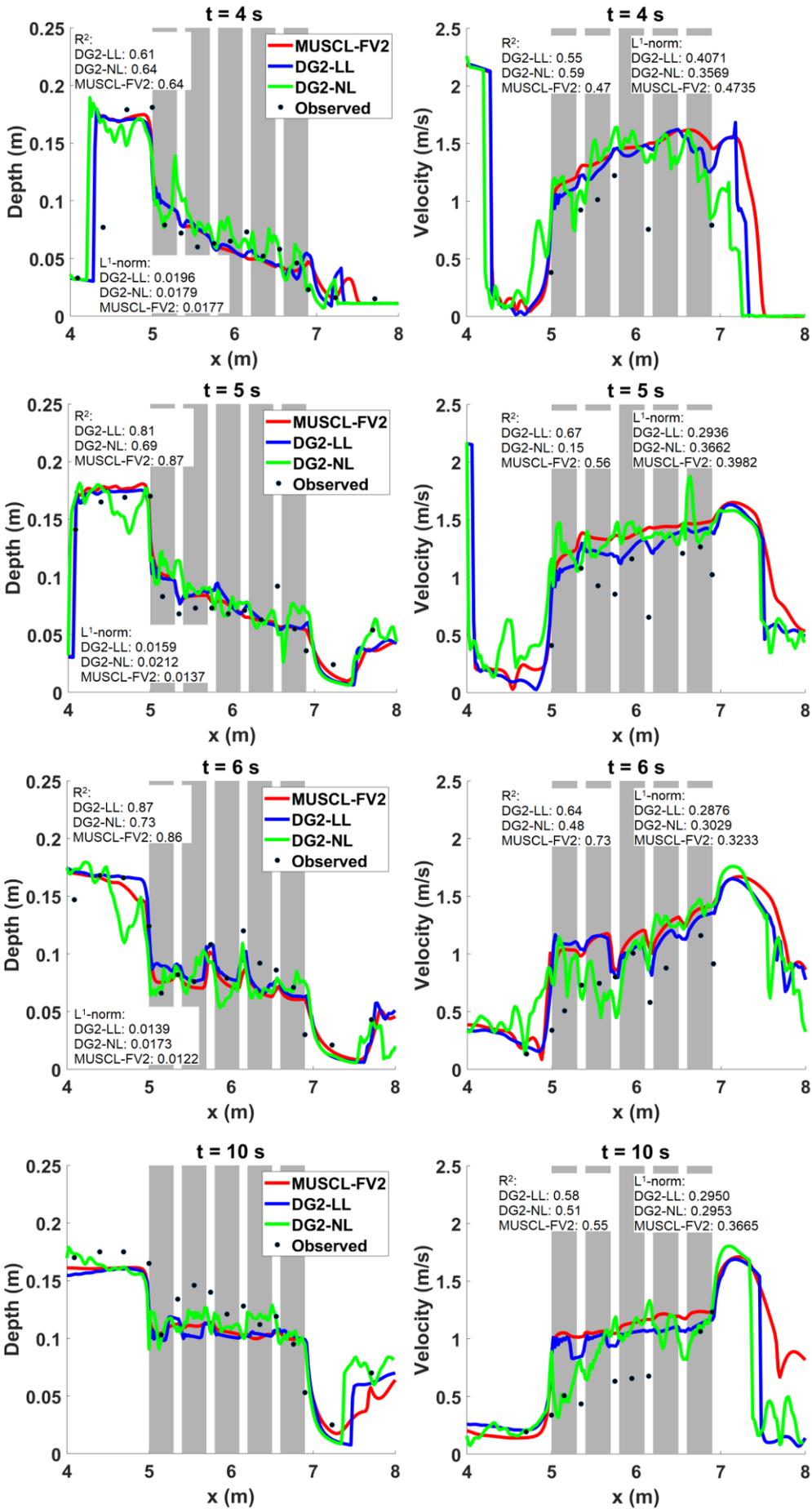

**Fig. 2.** Longitudinal profiles of the depths (left) and velocities (right) of MUSCL-FV2 (red line), DG2-LL (blue line) and DG2-NL (green line) extracted along $y = 0.2$ m relative to the experimental observations (black dots) at $t = 4, 5, 6$ and $10$ s. The grey shades indicate the building blocks, and the white strips between the blocks are the flow intersections. The $R^2$ and $L^1$-norm errors of each model are provided in the figure.

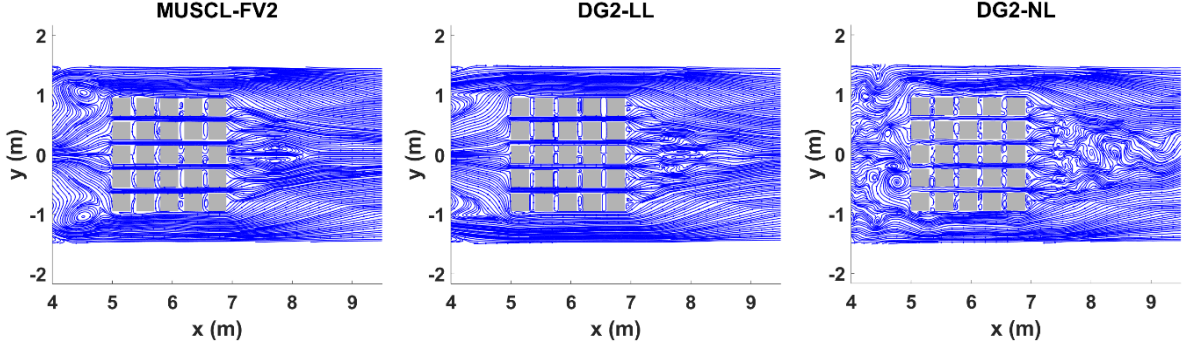

**Fig. 3.** Streamlines of 2D velocity fields produced by the MUSCL-FV2 (left), the DG2-LL (middle) and the DG2-NL (right) solvers around all blocks at $t = 10$ s.

As seen in Fig. 2, the DG2-LL produces relatively smooth patterns for the longitudinal profiles that are akin to those generated by MUSCL-FV2, but in a slightly better agreement with the observed profiles. In contrast, DG2-NL generates profiles with widespread noises, as expected when slope limiting is disabled within highly dynamic flow transitions around steep obstacles. Still, DG2-NL profiles are observed to trail the observed profiles reasonably well, in particular at the early stages of the dam-break wave propagation. This can be seen in the profiles predicted at $t = 4$ s, when the DG2-NL results are better correlated to the observed experimental data. As water becomes deeper within the gaps at $t = 5$ s, the DG2-LL retains a significantly stronger correlation, whereas the DG2-NL shows the most deviated prediction from the experimentally observed profiles, as the largest $L^1$-norm error indicates. At $t = 6$ s and $t = 10$ s, the DG2-LL also stands out in producing profiles that more closely resemble those of the MUSCL-FV2 (compare their $R^2$-error in Fig. 2) and are closer to the observed profiles compared to DG2-NL (compare their $L^1$-norm errors in Fig. 2). For the velocity profiles, in particular, DG2-LL gives the best fit of the velocity profiles to the observed profiles at the output time, $t = 10$ s, where it shows the highest $R^2$ and the lowest $L^1$-norm errors amongst all the models. The findings in Fig. 2 suggest that the DG2-LL is a better option to more reliably predict velocities in highly transient flows including complex interactions with topographic structures.

To gain deeper insights into the velocity patterns observed in the longitudinal profiles in Fig. 2, analyses have been performed considering the streamlines extracted from the numerically predicted 2D velocity fields around the blocks at the output $t = 10$ s. The streamlines made from the predictions by the MUSCL-FV2, the DG2-LL and the DG2-NL are shown in Fig. 3. The DG2-LL displayed smooth streamlines, consistent with its longitudinal profiles (Fig. 2), with regular distribution patterns around the blocks particularly in the upstream backwater zone ($x < 5$ m), in the *x*-directional gaps between the blocks (5 m $\leq x \leq$ 6.9 m) and in the wake downstream, where successive converging crossflows are noted with two elongated eddies at the wake edge ($x > 6.9$ m). Similar features are also seen in the MUSCL-FV2 streamlines, albeit with two distinct eddies formed in the backwater zone close to flume walls. However, the DG2-NL streamlines show chaotic distribution patterns with extensive spread of noises that are manifested as eddies in the backwater zone. These noises also lead to a very turbulent wake and forms swirling eddies at the wake edge resembling vortex shedding at $t = 10$ s. The analysis of the streamlines imply that the DG2-NL can still lead to misleading 2D velocity predictions and should be avoided when modelling 2D velocity fields. Besides, for this 10 s simulation, the DG2-NL is 3 times more expensive to run than the DG2-LL because of the need to adopt increasingly smaller time-steps (impacted by the noises). Instead, the DG2-LL seems a more reliable option for applications that require detailed capturing of 2D velocity fields, even with very smooth subcritical flow as demonstrated next.

### *Flow separation at a T-junction*

This test case involves mainstream subcritical flow separated into the lateral branch of a right-angled T-junction, and is run to examine further the performance of DG2-LL and DG2-NL in producing spatial 2D velocity fields and the lateral profiles from experimental observation (Bazin 2013; Bazin et al. 2017; Mignot et al. 2013). Two configurations are considered, as shown in Fig. 4: one without a 5 cm square obstacle located in the main branch, which represents a classical flow diversion into the lateral branch where the velocity field becomes more complex, and another with obstacle that would also affect the velocity field in the main branch. For both configurations, a steady inflow discharge of 0.002 m$^3$s$^{-1}$ is considered, associated with a Froude number of 0.23. entering the main branch from the eastern boundary. Fixed water depths of 45.1 mm and 44.6 mm from Bazin (2013) are imposed at the end of the downstream and the lateral branches, respectively. The bed surface is flat with a roughness value of $n_M = 0.01$ m$^{1/3}$s$^{-1}$, which is within the range of roughness values measured in the laboratory experiment. The simulations are run on a mesh with grid size of 0.005 m, which is the finest size used by the 2D-SWE models in Bazin et al. (2017), until steady-state was reached. Simulations are performed for each of the configurations using the MUSCL-FV2, DG2-LL and DG2-NL solvers, and analyses are conducted by considering the streamlines produced from the simulated 2D velocity fields, and by comparing the components of the simulated spatial velocity profiles with the available experimental data.

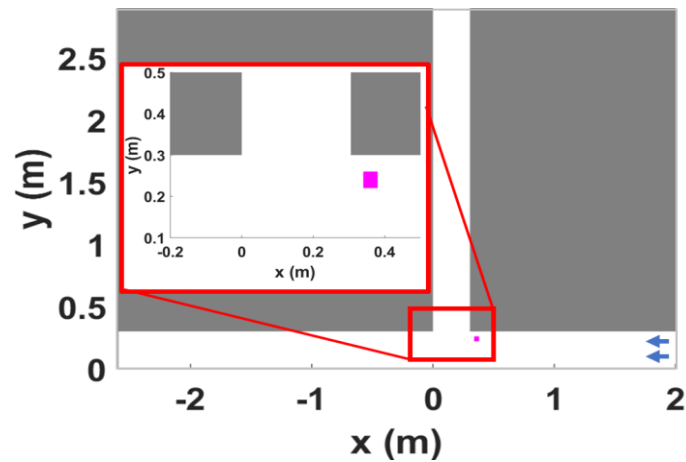

**Fig. 4.** T-junction spatial domain showing the main (white horizontal band) and the lateral (white vertical band) branches, and the upstream inflow (blue arrows) from the east of the main branch. The insert plot (red box) contains the zoomed-in view of the small obstacle position (magenta square) near the T-junction.

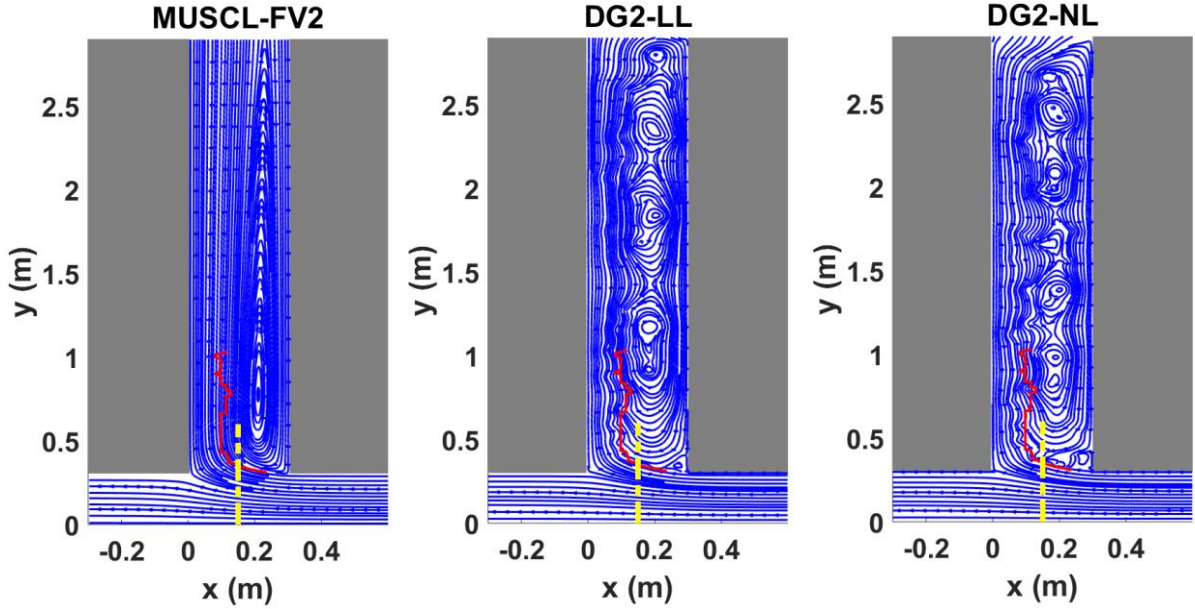

**Fig. 5.** Streamlines of the 2D velocity fields obtained with MUSCL-FV2 (left), DG2-LL (middle) and DG2-NL (right). In each subpanel, the measured recirculation extent in the lateral branch (red line) and the location of the extracted $u$-velocity and $v$-velocity at $x = 0.15$ m (yellow dashed line) for the case without the obstacle are also shown.

For the no-obstacle configuration, the extracted streamlines are shown in Fig. 5 indicates that, with all solvers, the mainstream flow is characterised by a region of separation along the eastern bank of the lateral branch. Within this region, MUSCL-FV2 streamlines exhibit a large flow recirculation centred close to the entrance of the lateral branch with regular, semi-elliptical patterns that narrows further downstream. In contrast, the DG2-LL and DG2-NL streamlines exhibit more uneven recirculation patterns, but resemble the measured flow recirculation pattern presented by Bazin et al. (2017), which is also quite uneven. Both the DG2-LL and DG2-NL streamlines also indicate a sequence of small-scale eddies within the inner zone of the recirculating flows, which is not observed with the MUSCL-FV2 streamlines. A major difference between the velocities predicted by DG2-LL and DG2-NL can, however, be observed for the $u$- and $v$-velocities extracted at the location $x = 0.15$ m near the entrance of the lateral branch. In Fig. 6, the corresponding velocity profiles are compared against the available experimental profiles of Bazin et al. (2017). The DG2-NL profiles exhibit sharp wavy patterns that are more apparent in the $v$-velocity profile, where a very sharp dip is predicted just after the entrance of the lateral branch ($0.3$ m $\leq y \leq 0.6$ m), suggesting the presence of a negative spurious velocity component. Such a deficiency can is confirmed by the fact that the DG2-NL generates the smallest $R^2$-error and the largest $L^1$-norm error, hence providing the least correlated and the most deviated predictions with respect to the experimental velocity profiles. In contrast, the profile of the DG2-LL velocity fields is much smoother and fits significantly better the experimental velocity data and those extracted from MUSCL-FV2. These analyses confirm that the DG2-LL solver is a better alternative to the DG2-NL to capture complex 2D velocity fields more reliably even when these fields occur under a subcritical flow regime with a low Froude number.

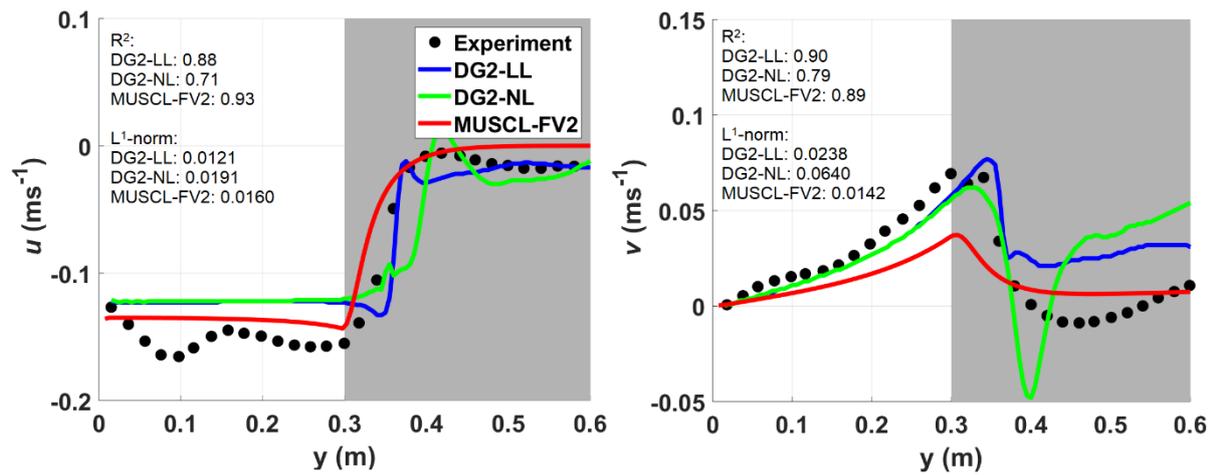

**Fig. 6.** Lateral profiles of the *u*-velocity (left) and *v*-velocity (right) from the MUSCL-FV2 (red line), the DG2-LL (blue line) and the DG2-NL (green line) solvers at *x* = 0.15 m, alongside the measured velocities (black dots) for the case without the obstacle. The $R^2$ and $L^1$-norm errors calculated along the lateral branch in each profile are also provided.

Similar analyses are presented for the results obtained from the MUSCL-FV2, the DG2-NL and the DG2-LL solvers using the configuration that included the obstacle upstream of the T-junction. The extracted streamlines are shown in Fig. 7, which clearly indicates more complex velocity fields behind the obstacle in the main branch. The MUSCL-FV2 streamlines exhibit a wake behind the obstacle with smooth recirculation extending downstream. This wake deflects the mainstream flow to the downstream part of the main branch, and thus effectively expands the recirculating flows in the lateral branch. The DG2-NL streamlines also display a wake behind the obstacle, but this is much smaller and cleared of any streamline, which suggests the presence of a "stagnant" zone (i.e. a zero velocity region). This smaller wake allows some of the mainstream flow to separate and form small-scale eddies within the lateral branch, which tend to progressively become more enlarged further downstream. In contrast, the DG2-LL streamlines do not show any wake behind the obstacle, suggesting that this solver can predict more mainstream flow separating into the lateral branch with a series of smaller-scale eddies that propagate along the eastern bank of the lateral branch.

Fig. 8 includes the lateral profiles of the $u$- and $v$-velocities predicted by the MUSCL-FV2, DG2-LL and DG2-NL solvers at $x = 0.15$ m. Because of the complexity of the flow patterns, neither the MUSCL-FV2 nor the DG2 solvers can produce profiles that capture the velocity profiles observed experimentally within the main branch (0 m ≤ $y$ ≤ 0.3 m). This means that the velocity fields in this region are too complex to be captured by the depth-averaged assumption featured in 2D SWE-based models without adding a turbulence closure or an eddy viscosity component (Bazin et al. 2017). Conversely, in the lateral branch, the $u$-velocity profiles from the MUSCL-FV2 and the DG2 solvers are comparatively more consistent with the experimental velocity data. Slight discrepancies are seen in the $v$-velocity profiles, in which DG2-NL yields the most deviated and least correlated predictions (compare $R^2$ errors and $L^1$-norm errors in Fig. 8). As supported further by the findings from Fig. 7 and Fig. 8, the DG2-LL is recommended for use when the aim of the application is focused on high-resolution analysis of spatial velocity fields that would be affected by the presence of sharp topographic edges, such as river groynes and building cavities.

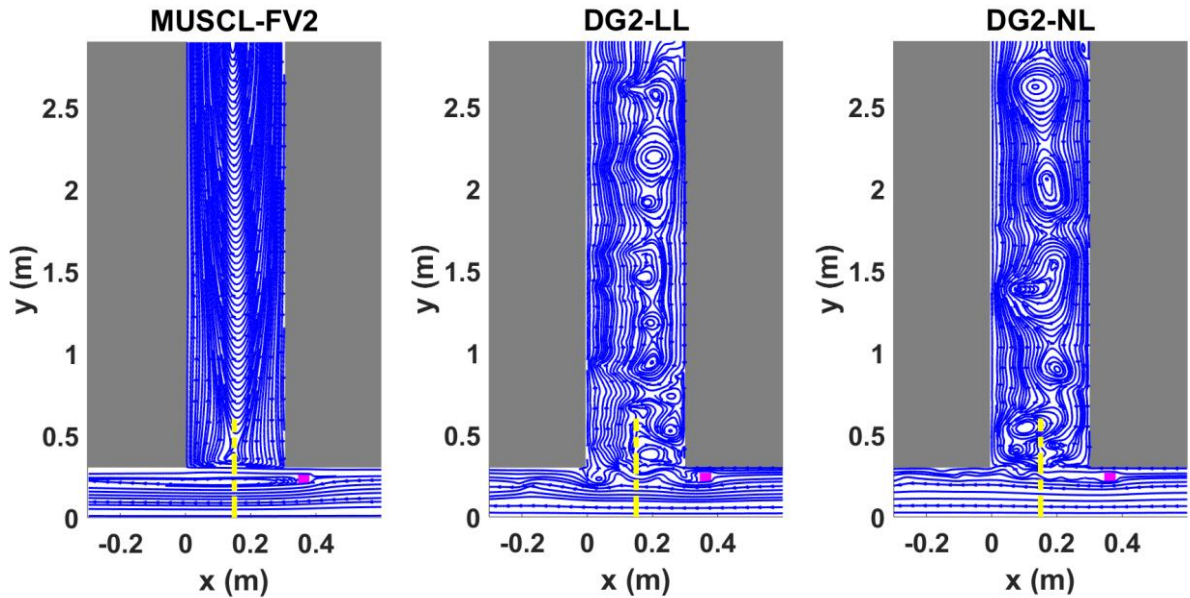

**Fig. 7.** Streamlines of the 2D velocity fields obtained with the MUSCL-FV2 (left), the DG2-LL (middle) and the DG2-NL (right) solvers for the case with the obstacle (magenta square). In each subpanel, the location of the extracted *u*-velocity and *v*-velocity at $x = 0.15$ m (yellow dashed line) is also shown.

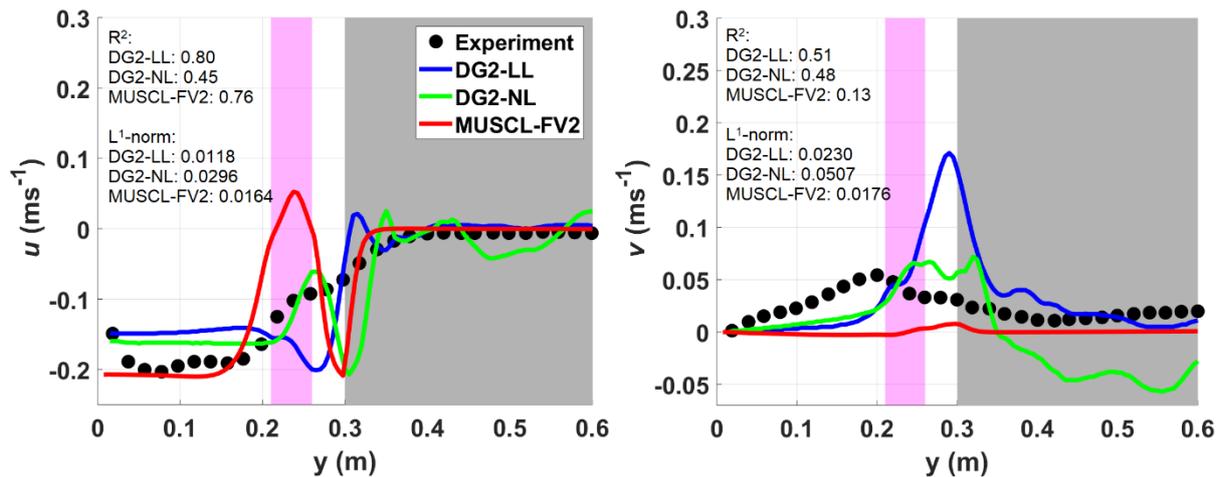

**Fig. 8.** Lateral profiles of the *u*-velocity (left) and *v*-velocity (right) for the MUSCL-FV2 (red line), the DG2-LL (blue line) and the DG2-NL (green line) solvers at $x = 0.15$ m, alongside the measured velocities (black dots) for the case with the obstacle (magenta square). The $R^2$ and $L^1$-norm errors calculated along the lateral branch in each profile are also provided in the figure.

*Quasi-steady flow in sharp building cavities*

In this test, the ability of the DG2-LL to produce more reliable spatial velocity fields is further examined by applying it to replicate high-resolution experimental data of 2D velocity fields. These data are obtained from particle image velocimetry (PIV) measurements performed in a physical model in the University of Sheffield involving one of the parking lots configurations with a closed manhole (Rubinato et al. 2021). The configuration is illustrated in Fig. 9, and installed on top of an existing experimental flume (Rubinato 2015). The flume slopes at 0.001 m/m and has a smooth surface ($n_M$ = 0.011 $m^{1/3}s^{-1}$), where a quasi-steady flow develops. The flow is driven by an inflow characterised by skewed velocities, which are extracted from the measured 2D velocity fields along the flume's southern boundary (red box, Fig. 9). At the northern boundary of the flume, free outflow conditions are imposed. DG2-LL and MUSCL-FV2 are run on a mesh with a grid size of 0.016 m, which matches the resolution of the PIV data. Simulated 2D velocity fields are extracted every 10 s and contrasted against the measured velocity fields based on the $L^1$-norm error.

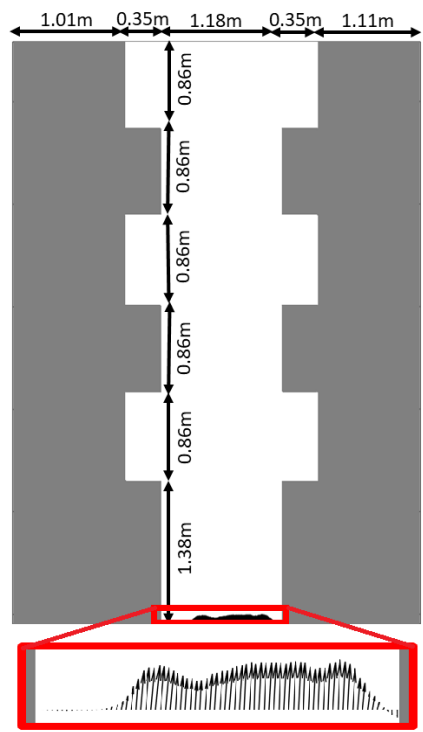

**Fig. 9.** The test configuration involving the large buildings (grey area) and side cavities. The black arrows in the red box indicate the inflow velocities extracted from the PIV-measured velocity fields along the southern boundary of the flume.

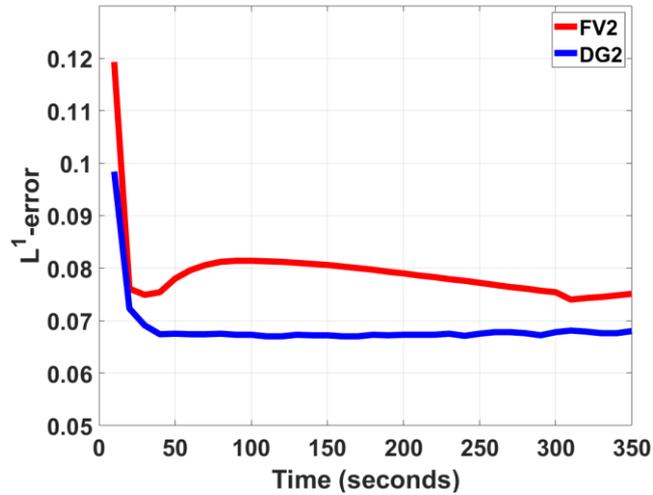

**Fig. 10.** The time history of $L^1$-norm errors based on the velocity fields produced by MUSCL-FV2 (red line) and DG2-LL (blue line) up to $t = 350$ s.

The time histories of the $L^1$-norm errors, up to 350 s, are illustrated in Fig. 10, for the DG2-LL and the MUSCL-FV2 solvers in order to analyse the runtime costs in producing the least deviated 2D velocity fields predictions. A sharp decrease in the $L^1$-norm errors is observed with both solvers in the first 20 s of the simulation. After then, a further decrease in the $L^1$-norm error is noted for the DG2-LL solver that stabilises from 40 s onwards, suggesting that any 2D velocity prediction by this solver has the least deviation from the measured velocity fields after 40 s. In contrast, the MUSCL-FV2 solver seems to require at least 300 s of simulation time for the $L^1$-norm error to stabilize at values that result slightly larger than for the DG2-LL. Overall, the $L^1$-norm error analysis indicates that in order to predict the closest velocity fields to the experimental data the MUSCL-FV2 solver should be around 7.5 times more expensive to run than the DG2-LL. Even at that cost, the MUSCL-FV2 results are not as accurate as the those of the DG2-LL, as discussed next.

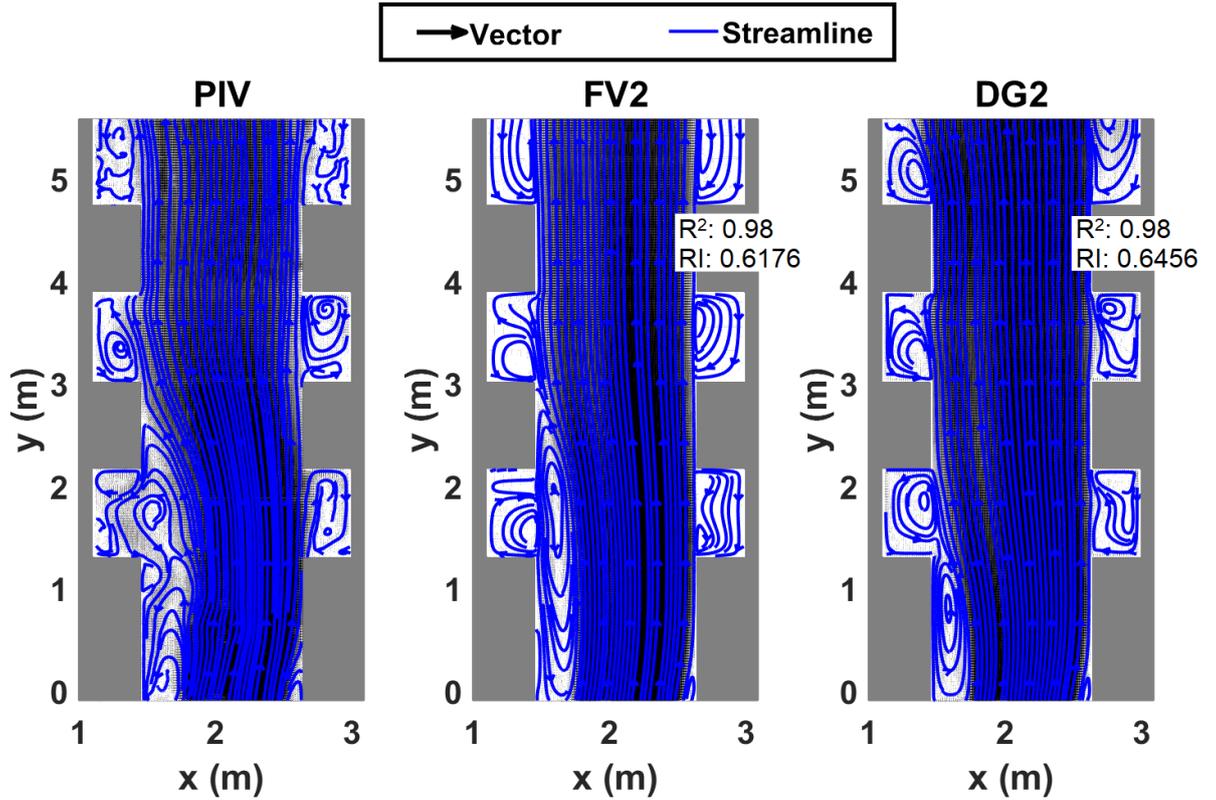

**Fig. 11.** Spatial vectors and streamlines of 2D velocity fields obtained from the PIV data (left), MUSCL-FV2 (middle) and DG2-LL (right) at $t = 350$ s with their respective $R^2$ and RI coefficients.

Fig. 11 displays the spatial vector map and its corresponding streamlines extracted from the 2D velocity fields experimentally measured using the PIV, relative to those predicted by the MUSCL-FV2 and the DG2-LL at $t = 350$ s. The spatial vector maps distinguish the regions of higher velocities, indicated by darker vector shades, to those with lower velocities, with much lighter vector shades. In the high velocity region, both the MUSCL-FV2- and DG2-LL streamlines show consistent linear patterns relatively similar to the measured velocity streamlines, particularly at the location of the mainstream flow. More dissimilar patterns are observed within the low velocity region. The DG2-LL solver predicts relatively well the location and the recirculation extent of the two eddies near the upstream cavity of the left building, which is instead slightly overlooked by the MUSCL-FV2 solver. Additionally, the DG2-LL can replicate the recirculation patterns of the eddies centred near the southern edge of the left middle cavity, and near the northern edge in the right middle cavity. The similarities between the streamline patterns of the DG2-LL and the PIV-measured velocities can be quantitatively confirmed by analysing the RI coefficient, instead of $R^2$, for which the solver yields a stronger fit in terms of both the magnitude and the directionality of the measured velocity fields. The results of this test case and the previous one highlight the benefits of applying the DG2-LL to produce more detailed and reliable velocity fields of quasi-steady flows and at a fraction of the computational cost.

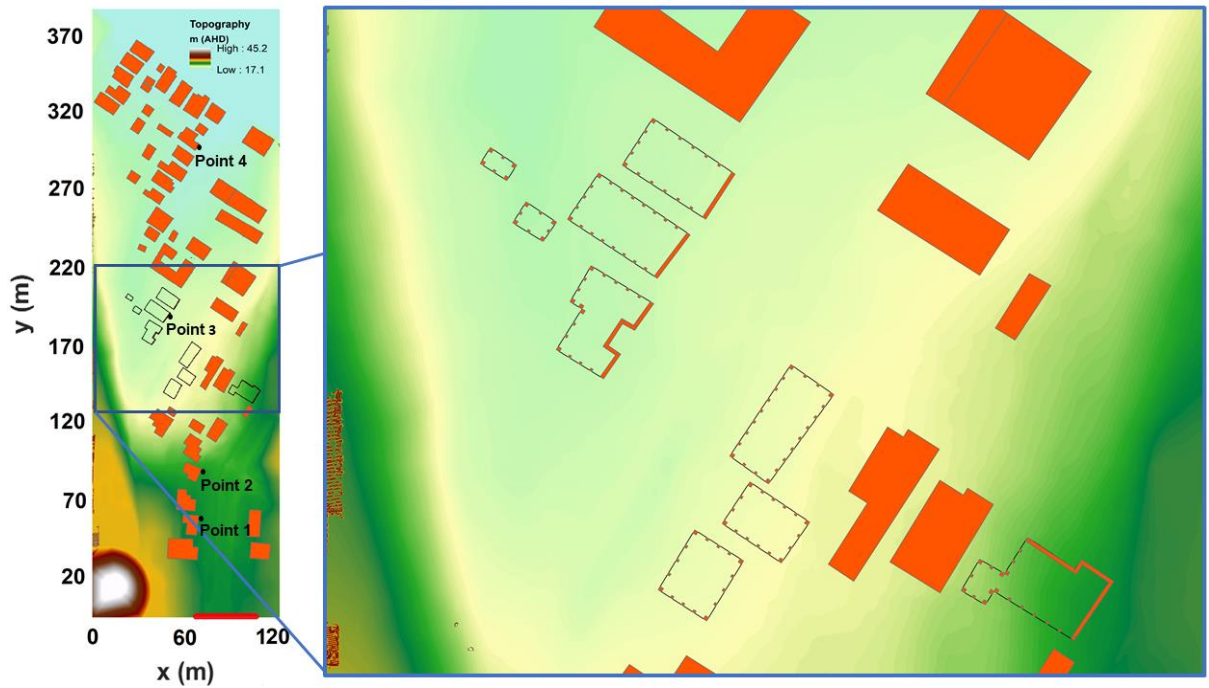

**Fig. 12.** The left panel illustrates the DEM of the Morgan-Selwyn floodway prototype, and includes the outlines of the buildings on piers (black boxes), water level sampling points (black dots) and the inflow boundary (red line, bottom of the panel). The blue box on the right provides a zoomed-in view of the walls and piers within the piered buildings (right panel).

### *Flooding in an urban residential area*

This test further investigates the DG2-LL capabilities to reproduce spatial velocity fields measured in a physical model for a flash flooding that spreads over an urban residential area with buildings on small piers. The physical model was developed by Smith et al. (2016) and consisted of a 12.5 m long and 5 m wide prototype representing the Morgan-Selwyn floodway in Merewether, Australia, to recreate the flooding scenario that occurred during the 'Pasha Bulker' storm. A steady-state inflow discharge of 19.7 $m^3$/s, which is the estimated peak flood discharge from the 'Pasha Bulker' storm, flows through a 38 m opening located at the southern boundary (red line, Fig. 12). This inflow propagates over the initially-dry topography that is constructed using road base materials. Free outflow conditions are imposed across the downstream boundary located at the north and wall boundary conditions are specified elsewhere. For the numerical model simulations, topographical survey points collated from the prototype are used to generate a digital elevation model (DEM) with a very fine grid size of 1 cm. This DEM is then resampled at a grid size of 17.5 cm (Ayog and Kesserwani 2021), in which square blocks with sizes of 52 cm for the corner piers and 35 cm for the side piers are added, as shown in Fig. 12. The dimensions of these piers are estimated by measuring the piers in the photographs provided in Smith et al. (2016). These piers are assumed to be unsubmerged along with other non-piered building blocks in the prototype. The flood simulations are run at the resampled DEM grid size up to $t = 800$ s at which steady state is reached at sampling point 4 located downstream of the piered buildings (left panel, Fig. 12).

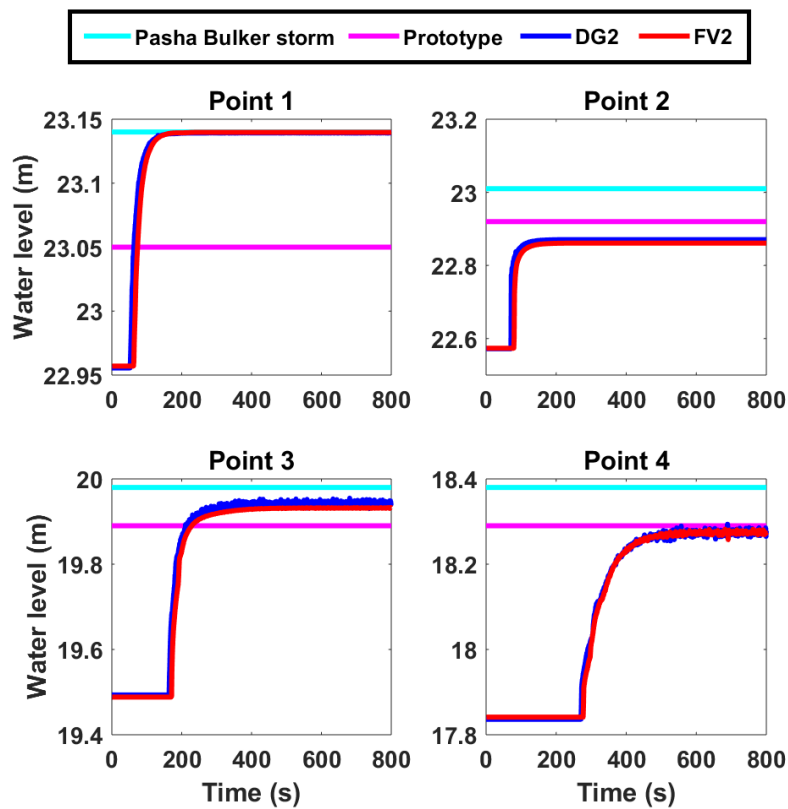

**Fig. 13.** The time histories of the water levels predicted by the DG2-LL and the MUSCL-FV2 solvers at all sampling points up to $t = 800$ s, alongside the historical flood levels records of the Pasha Bulker storm and the measured water level of the floodway prototype (Smith et al. 2016).

As recommended by Smith et al. (2016), the optimal roughness value, of $n_M = 0.042$ m$^{1/3}$s$^{-1}$, is calibrated by running both the MUSCL-FV2 and the DG2-LL solvers so as to closely reproduce the historical water level records of the 'Pasha Bulker' storm and the measured water levels of the prototype. Fig. 13 contains the associated time histories of the water level predicted by the MUSCL-FV2 and the DG2-LL solvers for this roughness value at the four sampling points shown in Fig. 12. It can be seen that the water level predictions by both solvers are in a good agreement with the measured water levels (maximum discrepancy of 0.1 m) and do not exceed the historical flood levels at point 1. To validate the velocity prediction around the piered buildings, velocity vectors produced by the MUSCL-FV2 and the DG2-LL solvers at $t = 800$ s were extracted at a 10 m spacing over the prototype where sparsely-measured velocity fields are available. Fig. 14 shows the velocity fields predicted by the solvers and the measured velocity vectors. Both solvers can effectively capture the sparsely-measured velocity fields within the region where the piered buildings are located. However, the DG2-LL solver yields slightly higher RI and lower L$^1$-norm error, indicating a better alignment with the measured velocity vectors.

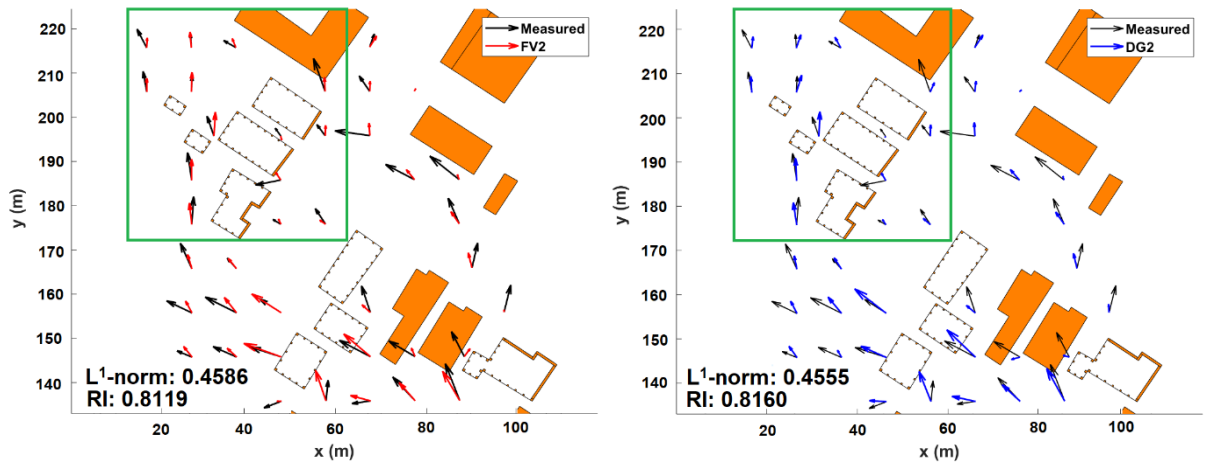

**Fig. 14.** The spatial vectors of the MUSCL-FV2 (left panel) and DG2-LL (right panel) velocity fields around the piered buildings at $t = 800$ s against the measured velocity vectors. The RI coefficient and L$^1$-norm errors are provided in each panel. The green boxes indicate the region where the streamlines of 2D velocity fields around the northwest piered buildings are analysed.

A more detailed investigation of the local distribution of the velocity fields around the piers is conducted by analysing the streamlines of the MUSCL-FV2 and the DG2-LL fine-resolution velocity predictions at $t$ = 800 s. Only zoomed-in portions are considered for the northwest piered buildings (green box, Fig. 14), where the flood wave interactions with the piers are more complex as shown in Fig. 15. The streamline plots indicate that the MUSCL-FV2 solver predicts chaotic recirculation patterns forming behind the front walls of buildings 1, 2 and 3. Similar recirculation patterns are also observed in the DG2-LL streamlines, but these are characterised by smaller-scale eddies coinciding with those observed previously with the DG2-LL in the flow separation region in the previous T-junction test. The DG2-LL also produces relatively irregular streamlines downstream of the back piers of buildings 2 and 3, which are less smooth than the MUSCL-FV2 streamlines. Additionally, the DG2-LL streamlines exhibit larger swirling patterns with eddies around piers southwest of building 1 and north of buildings 5 and 6. Hence, the analysis of the streamlines indicate that the DG2-LL solver can provide more precise velocity fields, and allows to gain a better insight on the localised flow patterns and (re)distributions around small topographical structures, such as when modelling flooding around piered buildings.

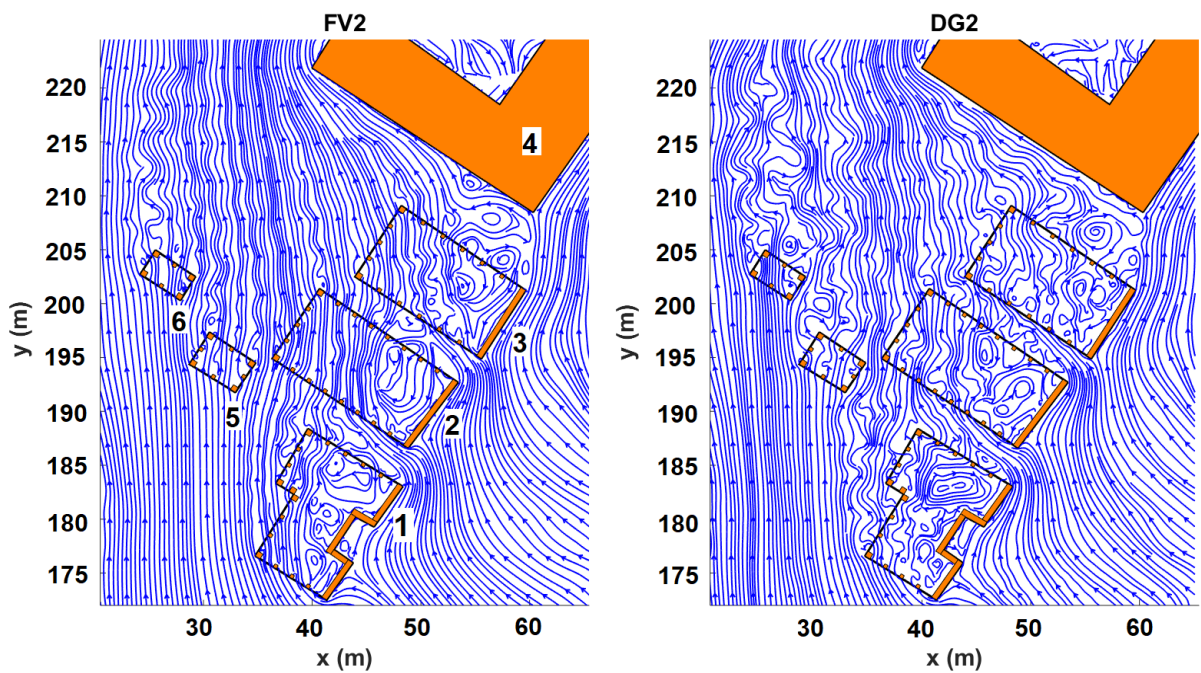

**Fig. 15.** The streamlines extracted from the MUSCL-FV2 (left panel) and the DG2-LL (right panel)

velocity fields around the northwest piered buildings at $t = 800$ s. The building numbers are provided in the left panel.

**Conclusions**

This paper has demonstrated the capabilities of a second-order discontinuous Galerkin (DG2) solver in reproducing detailed velocity fields when solving the depth-averaged shallow water equations without an eddy viscosity or turbulence term. These capabilities were investigated over four selected test cases for two DG2 variants with and without slope limiting, and by comparing their velocity predictions against those produced by the commonly used MUSCL-FV2 solver. The selected test cases required detailed capturing of small-scale velocity transients, at sub-meter scale, that occur at zones where the waves interact with steep topographies, for different flow conditions ranging from rapidly propagating transient flow to steady subcritical flow across a junction structure, within building side cavities, and behind small building piers. The solver predictions were compared using quantitative metrics and qualitative analysis of the predicted streamline at the zones of the recirculating flows. Performance analyses revealed that local limiting with the DG2 solver is needed to produce reliable velocity field predictions at sub-meter resolution even when the flow is subcritical and steady. Within this configuration, the DG2 solver is able to capture small-scale recirculation eddies in the velocity field prediction that are otherwise smeared out by the MUSCL-FV2 solver. Furthermore, the DG2 solver is more efficient than MUSCL-FV2, in the sense that it is much faster (around 7.5 times) to deliver the required steady state velocity predictions. These findings offer strong evidence that the DG2 solver is a valuable alternative to FV2 modelling, when detailed velocity fields are required within regions of wave-structure interactions, and without the need of additional mathematical complexity beyond standard shallow water equations.

**Data Availability Statement**

Some or all data, models, or code that support the findings of this study are available from the corresponding author upon reasonable request.

**Acknowledgements**

Georges Kesserwani acknowledges the support of the UK Engineering and Physical Sciences Research Council, grant ID: EP/R007349/1; and Janice Lynn Ayog acknowledges the financial support from the Malaysian Ministry of Education and Universiti Malaysia Sabah, Malaysia.